\documentclass[letterpaper,pra,floatfix,titlepage,twocolumn,aps,10pt,nofootinbib]{revtex4-1}
\pdfoutput=1

\usepackage{
	amssymb, % Math symbols such as \mathbb
	amsthm, % Theorem Formatting
	amsmath,
	mathtools,%this lets us define paired delimiters
	MnSymbol, %for \rrangle and \llangle
	physics,
	bm,
	ifthen,
	etoolbox, %for robust if/then control
	color, % Change text and page color
	appendix,
	phonetic,
	todonotes
} 

\newcommand{\draftmode}{1}    %to control draft colors below
\newcommand{\notetoself}[1]{\ifnum \draftmode=1 \todo[inline,  backgroundcolor=blue!20!white]{#1} \fi}
\newcommand{\cuttext}[1]{\ifnum \draftmode=1 \todo[inline,bordercolor=black!5!white,backgroundcolor=black!5!white]{\color{black!70!white} #1} \fi}
\newcommand{\todoj}[1]{\ifnum \draftmode=1 \todo[inline,backgroundcolor=green!15!white]{#1} \fi}
\newcommand{\warntext}[1]{\ifnum \draftmode=1 \todo[inline, bordercolor=orange!30!white, backgroundcolor=orange!30!white]{#1} \else #1 \fi}  
\let \projector \dyad % in package "physics", dyad{x}{y} makes |x><y| while dyad{x} makes |x><x|.  This just lets us call the second thing "\projector" too.

\newcommand{\avg}[1]{\left\langle#1\right\rangle} %<x>
\newcommand{\Dmat}{\mathbf{D}} %{D} %{\bar{D}} 
\newcommand{\Dmatin}{D} %{D} %{\bar{D}} 
\newcommand{\Ddiff}{D_0}%{\tilde{D}} %D^{\mathrm{diff}}} %diffusion coefficent\\
\newcommand{\DdiffDK}{D}%D_0
\newcommand{\Emat}{\mathbf{E}}
\newcommand{\Hmat}{\mathbf{H}} %{H} %Hamiltonian mat
\newcommand{\Hmatin}{H} %{H} %Hamiltonian mat
\newcommand{\lnVmatin}{Z}
\newcommand{\lnVmat}{\mathbf{Z}}
\newcommand{\Fmat}{\mathbf{K}} %{F} %{\bar{F}} %Forcing mat
\newcommand{\Fmatin}{K} %{F} %{\bar{F}} %Forcing mat
\newcommand{\Cmat}{\mathbf{C}}
\newcommand{\CVmat}{\mathbf{V}}

\newcommand{\CVTmat}{\overline{\mathbf{V}}} %{\tilde{\mathbf{V}}}
\newcommand{\CXmat}{\mathbf{Y}}%X}}
\newcommand{\CXTmat}{\overline{\mathbf{Y}}}%X}}}
\newcommand{\CXmatin}{Y}%X}
\newcommand{\CXTmatin}{\overline{Y}}%X}}
\newcommand{\Rmat}{\mathbf{R}}
\newcommand{\Amat}{\mathbf{B}}
\newcommand{\DisOp}{\hat{T}}
\newcommand{\alvec}{\boldsymbol{\alpha}}  %using boldsymbol instead of boldbf because the latter doesn't work on greek ketters
\newcommand{\betvec}{\boldsymbol{\beta}}
\newcommand{\muvec}{\boldsymbol{\mu}}
\newcommand{\Gfunc}{G}

\newcommand{\Id}{\mathbf{I}}
\newcommand{\IdT}{\overline{\mathbf{I}}}
\newcommand{\UCP}{\Omega}
\newcommand{\Wig}[1]{W_{#1}}%{W\{#1\}}
\newcommand{\Qig}[1]{Q_{#1}}%{Q\{#1\}}
\newcommand{\POVM}[2]{\Phi^{#1}_{#2}} %{\Phi^{#1}_{\overline{#2}}}
\newcommand{\sinv}{s}
\newcommand{\sreg}{s^{-1}}
\newcommand{\sregsq}{s^{-2}}
\newcommand{\jdet}[1]{\abs{#1}} %{\det #1}
\newcommand{\da}{\Omega} 
\newcommand{\dar}{\da_{\mathrm{R}}} 
\newcommand{\dai}{\da_{\mathrm{I}}}
\newcommand{\rr}{r}

\renewcommand{\draftmode}{0} %1 for drafts (faded text, etc.), 0 to not

\begin{document}
%%%%%%%%%%%%%%%%%%%%%%%%%%%%%

%%%%%%%%%%%%%%%%%%%%%%%%%%%%%
%      Title/authors        %
%%%%%%%%%%%%%%%%%%%%%%%%%%%%%

\title{Quantum Brownian motion as an iterated entanglement-breaking measurement by the environment}
\date{\today}
\author{C.~Jess~Riedel}\email{jessriedel@gmail.com}
\affiliation{Perimeter Institute for Theoretical Physics, Waterloo, Ontario N2L 2Y5, Canada}

%%%%%%%%%%%%%%%%%%%%%%%%%%%%%
%         Abstract          %
%%%%%%%%%%%%%%%%%%%%%%%%%%%%%

\begin{abstract}
Einstein-Smoluchowski diffusion, damped harmonic oscillations, and spatial decoherence are special cases of an elegant class of Markovian quantum Brownian motion models that is invariant under linear symplectic transformations.  Here we prove that for each member of this class there is a preferred timescale such that the dynamics, considered stroboscopically, can be rewritten exactly as unitary evolution interrupted periodically by an entanglement-breaking measurement with respect to a fixed overcomplete set of pure Gaussian states.  This is relevant to the continuing search for the best way to describe pointer states and pure decoherence in systems with continuous variables, and gives a concrete sense in which the decoherence can be said to arise from a complete measurement of the system by its environment.  We also extend some of the results of Di\'{o}si and Kiefer to the symplectic covariant formalism and compare them with the preferred timescales and Gaussian states associated with the POVM form.
\end{abstract}

\maketitle

%%%%%%%%%%%%%%%%%%%%%%%%%%%%%
%      Introduction         %
%%%%%%%%%%%%%%%%%%%%%%%%%%%%%

Although it has been widely studied for the better part of a century, the dynamical equations for Markovian quantum Brownian motion (QBM) \cite{lindblad1976brownian,dekker1981classical,isar1994open} were not solved in full symplectic generality until relatively recently \cite{brodier2004symplectic,wiseman2014quantum,robert2012time}.  By \emph{symplectic generality}, we mean a class of dynamical equations for a quantum state that is invariant under linear symplectic transformations of phase space, i.e., transformations of $(x,p)$ that are linear and preserve the symplectic form, in close analogy with Lorentz covariance.  Equation \eqref{eq:QBM-general} below describes the minimal such class of dynamics that subsumes the harmonic oscillator, frictionless spatial decoherence, and Einstein-Smoluchowski (frictionful, noninertial) diffusion. QBM models have been an essential testbed for understanding decoherence and the quantum-classical transition in systems with continuous degrees of freedom \cite{caldeira1983path, joos1985emergence,unruh1989reduction,zurek1993coherent,diosi2000robustness,hu1992quantum,strunz2002decoherence,dalvit2005predictability}, especially in the special case that can be generated by an environmental bath of oscillators coupled linearly to position.

In discrete systems, \emph{pure decoherence} \cite{zurek1981pointer,zwolak2014amplification} serves as a platonic ideal to which many real-world systems are close approximations.  Pure decoherence of a system occurs when the system's reduced dynamics take the form of a dephasing channel with respect to some orthonormal \emph{pointer basis} \cite{zeh1973toward,kubler1973dynamics,zurek1981pointer}.  However, the natural analogs of pure decoherence and the pointer basis remain elusive for systems with continuous degrees of freedom; several different definitions for the pointer basis in such cases have been proposed \cite{zurek1981pointer,zurek1993preferred,anglin1996decoherence,diosi2000robustness,dalvit2001unconditional,busse2010pointer,boixo2007generalized,yun2015analytic} but none are widely accepted.

It is now well appreciated that complete pure decoherence in discrete systems is equivalent to a complete measurement of the discrete variable by the environment.  In this article we generalize this to continuous systems. We prove that time-homogeneous Markovian QBM, when considered as a stroboscopic evolution between discrete times, is \emph{exactly} equivalent to unitary evolution punctuated periodically by a fixed positive operator-valued measurement (POVM) with respect to an overcomplete set of Gaussian wavepacket states.  In other words, the non-unitary component of QBM evolution is described by an \emph{entanglement-breaking} \cite{horodecki2003entanglement, wilde2013quantum} Gaussian measurement carried out by the environment.  This result can be straightforwardly extended to the case where the Hamiltonian and Lindblad operators are time-dependent \cite{ShuyiWriteup}.  This immediately recovers earlier work showing that, under QBM, the Wigner function becomes strictly and permanently positive in finite time for an arbitrary initial state \cite{diosi2002exact, eisert2004exact,brodier2004symplectic}.

It is tempting to suggest that the Gaussian wavepackets composing that POVM should reasonably be identified as the \emph{precise} pointer states of this open system evolution.  However, even when the dynamics are cast into the POVM form, there is still significant freedom to choose the preferred pointer states. The choices are close to, but generally distinct from, the (also Gaussian) pointer states suggested by Di\'{o}si and Kiefer \cite{diosi2000robustness, diosi2002exact} and by Zurek and collaborators \cite{zurek1993preferred,dalvit2005predictability}.  Such alternatives are related to different criteria of classicality that have been studied in the past, such as the Wigner function becoming positive \cite{diosi2002exact,brodier2004symplectic,eisert2004exact},  the Glauber $P$ function becoming positive \cite{diosi1987exact, diosi2002exact} (and no more singular than a $\delta$ function \cite{genoni2013detecting}), or that the quantum state $\rho$ is expressible as an incoherent mixture of Gaussian states \cite{genoni2013detecting}.  Thus, our results add to the zoo of possible classicality criteria and associated pointer states used to understand decoherence in QBM, although most sensible choices are closely related for dimensional reasons.

In Section I we state the main result after introducing the minimum necessary notation, and in Section II we give a proof.  In Section III we connect this to other preferred states and timescales discussed in the literature, especially to the work of Di\'{o}si and Kiefer. In Section IV we offer concluding discussion.  A summary of symplectic QBM using our notation can be found in Appendix A, including a description of important special cases.  In Appendix B we explicitly compute the POVM form of the dynamics for the common special case of a lightly damped harmonic oscillator.

%%%%%%%%%%%%%%%%%%%%%%%%%%%%%
\section{Main result}
%%%%%%%%%%%%%%%%%%%%%%%%%%%%%

Markovian QBM in the symplectic general form is given by the master equation $\partial_t \rho_t = \mathcal{L} [\rho_t]$ where the manifestly covariant Linblad superoperator is
\begin{align}
\label{eq:QBM-general}
\mathcal{L} [\rho_t] \equiv - i \frac{\Fmatin_{ab}}{2\hbar}   \left[\hat{\alpha}^a,\left\{ \hat{\alpha}^b, \rho_t \right\} \right] - \frac{\Dmatin_{ab}}{2 \hbar^2} \left[ \hat{\alpha}^a,\left[\hat{\alpha}^b, \rho_t \right] \right],
\end{align}
for the density matrix $\rho_t$ of a single continuous degree of freedom \cite{lindblad1976brownian,dekker1981classical,isar1994open}. Here, $\hat{\alpha} = (\hat{x},\hat{p})$ is a vector operator for a point in phase space, repeated indices are summed over the two phase-space directions ($a = x,p$), $\Dmatin_{ab}$ is a positive semidefinite 2-by-2 matrix with real entries, and $\Fmatin_{ab}$ is a 2-by-2 matrix with real entries satisfying $(\Fmatin^a_{\phantom{a}a})^2 \le 2 D_{ab}D^{ab}$.   Indices are raised and lowered with the symplectic form using the anti-symmetric Levi-Civita tensor $\epsilon^{ab}$, e.g., $\Fmatin^a_{\phantom{a}b} = \epsilon^{ac} \Fmatin_{cb}$.  (They behave just like Weyl spinors.)

We set $\hbar = 1$ and introduce notation where $\Fmatin^a_{\phantom{a}b}$, $D^{ab}$, and $\alpha^a$ are replaced by boldface $\Fmat$, $\Dmat$, and $\alvec$.  We define $\gamma \equiv - \Tr \Fmat/2 = - \Fmatin^a_{\phantom{a}a}/2 = \Fmatin_a^{\phantom{a}a}/2$ and $\Hmat \equiv \Fmat + \gamma \Id$ so that $\Hmat$ is traceless and $0 \le  \gamma ^2 \le \jdet{\Dmat}$, where $\jdet{\,\cdot\,}$ denotes the determinant.  Let $\CVmat$ be an arbitrary matrix with unit determinant representing a canonical linear transformation, i.e., a (classical) linear transformations on phase space which preserves the symplectic form.  We denote the associated quantum unitary evolution by $\hat{U}_{\CVmat}$, so that $\avg{\hat{U}_{\CVmat}^\dagger \hat{\alvec}\hat{U}_{\CVmat}} = \CVmat \avg{\hat{\alvec}}$ and $\hat{U}_{\CVmat}^\dagger = \hat{U}_{\CVmat^{-1}}$.  The corresponding superoperator on the space of density matrices is $\UCP_\CVmat[\rho] \equiv \hat{U}_\CVmat \rho \hat{U}_\CVmat^\dagger$.  Let $\ket{\alvec}$ represent the normalized coherent states with phase-space mean $\alvec =(x,p)$, so that $\ket{\CVTmat;\alvec} \equiv \hat{U}_\CVmat \ket{\CVmat^{-1} \alvec}$ are the pure Gaussian states \cite{combescure2012coherent} parametrized by their mean $\alvec \equiv \avg{\hat{\alvec}}$ and their 2-by-2 covariance matrix $\CVTmat$ where $\CVTmat \equiv \CVmat \CVmat^\intercal/2$.

Lastly, for $s>0$,
\begin{align}
\POVM{\sinv}{\CVTmat} [\rho] \equiv \int \! \frac{\dd{\alvec}}{2\pi} \matrixelement{\CVTmat;\alvec}{\rho}{\CVTmat;\alvec} \cdot \ketbra{\CVTmat; \sinv \alvec}{\CVTmat; \sinv \alvec} 
\end{align}
represents the entanglement-breaking channel \footnote{A channel $\Phi_\mathcal{A}$ on a Hilbert space $\mathcal{A}$ is defined to be \emph{entanglement breaking} if it always produces separable states when operating on one part of an entangled pair, that is, when $(\Phi_{\mathcal{A}} \otimes I_{\mathcal{B}})[\projector{\Psi_{\mathcal{AB}}}]$ is separable for all $\Psi_{\mathcal{AB}}$.  This is true if and only if all its Krauss operators are unit rank \cite{wilde2013quantum}.} given by the Krauss operators $\hat{A}_{\alvec} \propto \ketbra{\CVTmat; \sinv \alvec}{\CVTmat; \alvec}$.  This is a POVM measurement with respect to the overcomplete basis $\ket{\CVTmat; \alvec}$ -- formally, a \emph{frame} \cite{christensen2003introduction} -- where the  measurement outcome $\alvec$ is followed by a preparation of the state $\ketbra{\CVTmat; \sinv \alvec}$ (and then forgotten).  A minimally disturbing measurement corresponds to $s=1$, while dilations and contractions of phase space are given by $s>1$ and $s<1$, respectively.

\textbf{Theorem.} There is a characteristic time $T>0$ defined as the unique positive solution to $\jdet{\Cmat_T} = (1+e^{-2\gamma T})^2/4$, where
\begin{align}
\label{eq:POVM-time}
\Cmat_t \equiv \int_0^t\! e^{\tau \Fmat} \Dmat e^{\tau \Fmat^\intercal}  \dd{\tau}.
\end{align} 
Moreover, integrating the dynamics \eqref{eq:QBM-general} forward by $T$ induces a completely positive (CP) trace-preserving map that sequentially evolves, measures, and prepares the system,
\begin{align}
\label{eq:POVM-equation}
%e^{T\mathcal{L}} &= \POVM{e^{-\gamma T}}{2\Cmat_T^{1/2}/(1-e^{-2 \gamma T})} \circ \UCP_{ e^{T\Hmat}}
e^{T\mathcal{L}} = \POVM{\sinv}{\CVTmat} \circ \UCP_{\Rmat}
\end{align}
where $\circ$ denotes composition, and where $\CVTmat =\Cmat_T/(1+e^{-2\gamma T})$, $s = e^{-\gamma T}$, and $\Rmat = e^{T \Hmat}$.

\textbf{Remark.} This can be rewritten in several ways:
\begin{align}
\begin{split}
\label{eq:POVM-equation-rewrite}
e^{T\mathcal{L}} [\rho] &= \POVM{\sinv}{\CVTmat} \circ \UCP_{  \Rmat} [\rho ] \\
&= \UCP_{\CVmat} \circ \POVM{\sinv}{\IdT} \circ \UCP_{\CVmat^{-1} \Rmat} [\rho ] \\ 
&= \UCP_{\Rmat} \circ \POVM{\sinv}{\overline{\Rmat^{-1} \CVmat}}  [\rho ] \\
&= \int \! \frac{\dd{\alvec}}{2\pi} \matrixelement{\overline{\Rmat^{-1}\CVmat}; \alvec}{\,\rho\,}{\overline{\Rmat^{-1}\CVmat};\alvec} \\
& \quad \quad \times \projector{\CVTmat; \sinv \Rmat \alvec},
\end{split}
\end{align}
or generally as
\begin{align}
\label{eq:POVM-general}
e^{T\mathcal{L}} [\rho] = \UCP_{\Rmat^{1-\rr}} \circ \POVM{\sinv}{\CVTmat_{\rr}} \circ \UCP_{\Rmat^r} [\rho ].
\end{align}
where
\begin{align}
\begin{split}
\label{eq:POVM-general-mat}
%\CVTmat_{\rr}  &=  e^{-(1-\rr)T\Hmat} \frac{\Cmat_T}{1+s^2} e^{-(1-\rr)T\Hmat^\intercal} \\
\CVTmat_{\rr}  &\equiv  e^{-(1-\rr)T\Hmat} \frac{\Cmat_T}{1+e^{-2 \gamma T}} e^{-(1-\rr)T\Hmat^\intercal} \\
&= \Rmat^{-(1-\rr)} \CVTmat (\Rmat^\intercal)^{-(1-\rr)} \\
&= \overline{\Rmat^{-(1-\rr)} \CVmat}
%&=  e^{-(1-\rr)T\Hmat} \CXTmat e^{-(1-\rr)T\Hmat^\intercal}
\end{split}
\end{align}
for any $\rr \in [0,1]$, with special cases $\CVTmat_{0} =  \overline{\Rmat^{-1} \CVmat}$ and $\CVTmat_{1} =  \CVTmat$. Since the dynamics are Markovian, this can be iterated, e.g.,
\begin{align}
%e^{2T\mathcal{L}} = \POVM{e^{-\gamma T}}{\Cmat_T^{1/2}} \circ \UCP_{\Cmat_T^{-1/2} e^{T\Hmat}} \circ \POVM{e^{-\gamma T}}{\Cmat_T^{1/2}} \circ \UCP_{\Cmat_T^{-1/2} e^{T\Hmat}},
e^{2T\mathcal{L}} = \POVM{\sinv}{\CVTmat} \circ \UCP_{  \Rmat} \circ \POVM{\sinv}{\CVTmat} \circ \UCP_{  \Rmat},
\end{align}
and so on for $e^{nT\mathcal{L}}$ with $n$ any positive integer. In other words, quantum Brownian motion can be understood stroboscopically as an iterated phase-space measurement interspersed with unitary evolution.

%%%%%%%%%%%%%%%%%%%%%%%%%%%%%
\section{Proof}
%%%%%%%%%%%%%%%%%%%%%%%%%%%%%

Under \eqref{eq:QBM-general}, the Wigner function $\Wig{\rho}$ corresponding to the state $\rho_t$ is known to obey a Klein-Kramers dynamical equation \cite{isar1994open,brodier2004symplectic,wiseman2014quantum,robert2012time},
which can be written in sympectic covariant form as
\begin{align}
\label{eq:symplectic-wig-decoh}
\partial_t \Wig{\rho} (\alpha) = \left[ -\Fmatin^{a}_{\phantom{a}b} \partial_a \alpha^b + \frac{1}{2} \Dmatin^{ab} \partial_a \partial_b \right] W_{\rho}(\alpha) ,
\end{align}
where $\partial_a = \partial/\partial \alpha^a$.  Below we will work within the space of functions over phase space using the convolution operator $\ast$ and the composition operator $\circ$.  In this context, matrices are taken to represent the functions obtained by matrix multiplication with the phase space point $\alvec$, i.e., $ \CVmat (\alvec) = \CVmat \alvec$ and $(W_\rho \circ \CVmat)(\alvec) = W_\rho( \CVmat \alvec)$.

In this notation, the exact solution to \eqref{eq:wig-decoh} for any initial Wigner distribution $\Wig{\rho}(\alvec)$ is known to be \cite{brodier2004symplectic,wiseman2014quantum, robert2012time} 
\begin{align}
\label{QBM-solution}
\Wig{e^{t \mathcal{L}}[\rho]} =  \Gfunc_{\Cmat_t} \ast \left(e^{ 2 \gamma t}  \Wig{\rho} \circ e^{-t \Fmat} \right) 
\end{align}
where $\Cmat_t$ is given by \eqref{eq:POVM-time} and 
\begin{align}
\Gfunc_{\Amat}(\alvec) \equiv \frac{e^{-\frac{1}{2}\alvec^\intercal \Amat^{-1} \alvec}}{2\pi \sqrt{ \jdet{\Amat}}}
\end{align}
is a normalized Gaussian smoothing kernel for any positive semidefinite covariance matrix $\Amat$.  Note the important special case of unitary evolution, $\gamma = 0 = \Dmat$, for which $\Wig{e^{t\mathcal{L}}[\rho]} = \Wig{\rho} \circ e^{-t \Hmat}$.

The key idea in the proof is that the POVM-and-prepare channel $\POVM{1}{\IdT}$ with respect to the coherent states $\ket{\alvec} = \ket{\IdT;\alvec}$ corresponds, in the Wigner representation, to a convolution with the kernel $\Gfunc_{2\IdT} = \Gfunc_{\Id}$.  That is, $\Wig{\POVM{1}{\IdT}[\rho]} = \Gfunc_{2\IdT} \ast \Wig{\rho}$.  
More generally, dissipation in phase space can be accounted for by considering the modified channel 
\begin{align}
\POVM{\sinv}{\IdT}[\rho] = \int \! \frac{\dd{\alvec}}{2\pi}  \ketbra{\sinv \alvec}{\alvec} \rho \ketbra{\alvec}{\sinv \alvec}
\end{align}
and calculating the corresponding Husimi Q function
\begin{align}
\begin{split}
\Qig{\POVM{\sinv}{\IdT}[\rho]} (\alvec) &= \frac{1}{2\pi} \matrixelement{\alvec}{\POVM{\sinv}{\IdT}[\rho]}{\alvec} \\
&= \frac{1}{2\pi} \int \! \frac{\dd{\betvec}}{2\pi}  \matrixelement{\betvec}{\rho}{\betvec} \braket{\alvec}{\sinv \betvec} \braket{\sinv \betvec}{\alvec}\\
&= \int \! \frac{\dd{\betvec}}{2\pi}  \Qig{\rho}(\betvec) \abs{\braket{\alvec}{\sinv \betvec}}^2 \\
&= \int \! \dd{\boldsymbol{\mu}} \sregsq \Qig{\rho}(\sreg \boldsymbol{\mu}) \frac{\abs{\braket{\alvec}{\boldsymbol{\mu}}}^2}{2\pi} 
\end{split}
\end{align}
where we have changed integration variables to $\muvec = \sinv \betvec$. Using the fact that $\abs{\braket{\alvec}{\muvec}}^2 = \exp[-(\alvec-\muvec)^2/2] = 2\pi  G_{\Id}(\alvec - \muvec)$ we obtain
\begin{align}
\label{Q-func-scale}
\Qig{\POVM{\sinv}{\IdT}[\rho]} = \Gfunc_{2\IdT} \ast (\sregsq \Qig{\rho} \circ \sreg\Id )
\end{align}

The Husimi Q function is just the Wigner function smoothed by a Gaussian, $Q =  \Gfunc_{\IdT} \ast W$.  Using these directly checkable identities
\begin{gather}
\label{eq:id1}
(X \ast Y)\circ \mathbf{A} = \jdet {\mathbf{A}} \cdot (X \circ \mathbf{A}) \ast (Y \circ \mathbf{A}), \\
\Gfunc_{\mathbf{B}} \circ \mathbf{A}^{-1} = \jdet{\mathbf{A}} \cdot \Gfunc_{\mathbf{A} \mathbf{B} \mathbf{A}^\intercal}, \\
\label{eq:id3}
\Gfunc_{\mathbf{A}} \ast \Gfunc_{\mathbf{B}} = \Gfunc_{\mathbf{A}+\mathbf{B}},
\end{gather}
we can deconvolve both sides of \eqref{Q-func-scale} to get
\begin{align}
\Wig{\POVM{\sinv}{\IdT}[\rho]} &=\Gfunc_{(1+\sinv^2)\IdT} \ast \left(\sregsq  \Wig{\rho} \circ \sreg\Id \right)
\end{align}

This can be generalized to a POVM of Gaussian states $\ket{\CVTmat;\alvec} = \hat{U}_\CVmat \ket{\IdT; \CVmat^{-1} \alvec}$ with any linear symplectic transformation $\CVmat$ by first applying an appropriate unitary $\UCP_{\CVmat^{-1}}$, executing the measurement, and then applying the inverse unitary $\UCP_\CVmat$:
$\POVM{\sinv}{\CVTmat}[\rho] = \UCP_\CVmat \circ \POVM{\sinv}{\IdT} \circ \UCP_{\CVmat^{-1}} [ \rho ]$.
In the Wigner representation this is
\begin{align}
\begin{split}
\Wig{\POVM{\sinv}{\CVTmat}[\rho]} &= \Wig{\UCP_\CVmat \circ \POVM{\sinv}{\IdT} \circ \UCP_{\CVmat^{-1}} [\rho]} \\
&= \Wig{\POVM{\sinv}{\IdT}  \circ \UCP_{\CVmat^{-1}} [\rho]} \circ \CVmat^{-1} \\
&= \left( \Gfunc_{\left(1+\sinv^2\right)\IdT} \ast \left(\sregsq \Wig{\UCP_{\CVmat^{-1}}[\rho]} \circ \sreg\Id\right)\right) \circ \CVmat^{-1} \\
&= \left( \Gfunc_{\left(1+\sinv^2\right)\IdT} \ast \left(\sregsq \Wig{\rho} \circ \CVmat \circ \sreg\Id\right)\right) \circ \CVmat^{-1} \\
&= \left( \Gfunc_{\left(1+\sinv^2\right)\IdT} \circ \CVmat^{-1}\right) \ast \left(\sregsq \Wig{\rho} \circ \sreg\Id\right) \\
&=\Gfunc_{\left(1+\sinv^2\right)\CVTmat} \ast \left(\sregsq \Wig{\rho} \circ \sreg\Id\right).
\end{split}
\end{align}

Augmenting this with unitary evolution  $\UCP_\Rmat$ corresponding to an arbitrary linear symplectic transformation $\Rmat$ gives
\begin{align}
\begin{split}
%\Wig{ \POVM{\sinv}{\CVTmat} \circ \UCP_\Rmat[\rho]} 
%&=\Gfunc_{\left(1+\sinv^2\right)\CVTmat} \ast \left(\sregsq  \Wig{\UCP_\Rmat[\rho]} \circ \sreg \Id\right) \\
%&= \Gfunc_{\left(1+\sinv^2\right)\CVTmat} \ast \left(\sregsq \Wig{\rho} \circ \sreg \Rmat^{-1}\right)
\Wig{ \POVM{\sinv}{\CVTmat} \circ \UCP_\Rmat[\rho]} 
= \Gfunc_{\left(1+\sinv^2\right)\CVTmat} \ast \left(\sregsq \Wig{\rho} \circ \sreg \Rmat^{-1}\right)
\end{split}
\end{align}

We can then reproduce the solution \eqref{QBM-solution} by choosing $\CVTmat =\Cmat_T/(1+\sinv^2)$, $s=e^{-\gamma T}$, and $\Rmat = e^{\Hmat T}$.  However, this is only possible when $\jdet{\Cmat_T} = (1+\sinv^2)^2/4$, since we have assumed $\CVmat$ is a linear symplectic transformation (so $\jdet {\CVTmat} = \jdet{\CVmat}^2/4 = 1/4$).  Let us prove that this requirement uniquely determines $T$ so long as $\gamma$ is not extremal (i.e., so long as $\gamma^2 < \jdet{\Dmat}$ rather than $\gamma^2 = \jdet{\Dmat}$).

First, note that since $\Dmat$ is positive semidefinite, the integrand of \eqref{eq:POVM-time} is also positive semidefinite by construction. By the Minkowski determinant theorem (see, for example, Ref.\ \cite{marcus1992survey}), the determinant function obeys $\sqrt{\jdet{\mathbf{A}+\mathbf{B}}} \ge \sqrt{\jdet {\mathbf{A}}} + \sqrt{\jdet {\mathbf{B}}}$ for positive semidefinite 2-by-2 matrices, so 
\begin{align}
\begin{split}
\label{eq:det-c-bound}
\jdet{\Cmat_t} &\ge  \left[\int_0^t\! \sqrt{ \jdet{e^{t\Fmat} \Dmat e^{t\Fmat^\intercal}} } \dd{\tau}\right]^2 \\
&=  \left[  \int_0^t\! \left(e^{\tau \Tr \Fmat}\right) \dd{\tau}\right]^2 \jdet {\Dmat}  \\
&=  (1-e^{-2 \gamma t})^2 \frac{\jdet{ \Dmat}}{4\gamma^2},
\end{split}
\end{align} 
and likewise
\begin{align}
\begin{split}
\label{eq:det-c-d-t-bound}
\frac{\dd \jdet{\Cmat_t}}{\dd t} &\ge 2 e^{-2 \gamma t} \sqrt{\jdet{\Dmat} \jdet{\Cmat_t}}.
%\frac{4 \jdet{\Cmat_t}}{(1+e^{-2 \gamma t})^2} &\ge \frac{4}{(1+e^{-2 \gamma t})^2} \left[\int_0^t\! \sqrt{ \jdet{e^{t\Fmat} \Dmat e^{t\Fmat^\intercal}} } \dd{\tau}\right]^2 \\
%&=  \frac{4}{(1+e^{-2 \gamma t})^2}  \left[  \int_0^t\! \left(e^{\tau \Tr \Fmat}\right) \dd{\tau}\right]^2 \jdet {\Dmat}  \\
%&=  \tanh^2 (\gamma t) \frac{\jdet{ \Dmat}}{\gamma^2}.
\end{split}
\end{align} 
From this one can show that $4 \jdet{\Cmat_t}/(1+e^{-2 \gamma t})^2$ increases monotonically with $t$ and moreover
\begin{align}
\begin{split}
\label{eq:mod-det-c-bound}
\frac{4 \jdet{\Cmat_t}}{(1+e^{-2 \gamma t})^2} &\ge  \tanh^2 (\gamma t) \frac{\jdet{ \Dmat}}{\gamma^2}.
\end{split}
\end{align} 
For $\gamma \neq 0$, the function $\tanh^2 (\gamma t)$ starts at zero and approaches unity as $t \to \infty$. Thus for $\jdet {\Dmat} \neq 0$, we have that $4 \jdet{\Cmat_T}/(1+e^{-2 \gamma T})^2 = 1$ for some unique finite $T$, except for the extremal cases $\gamma = \pm \sqrt{\jdet{ \Dmat}}$ (including $\Dmat=0$) for which $T \to \infty$.

%%%%%%%%%%%%%%%%%%%%%%%%%%%%%
\section{Other states and timescales}
%%%%%%%%%%%%%%%%%%%%%%%%%%%%%

In this section, we consider the preferred timescales and pointer states discussed by Di\'{o}si and Kiefer \cite{diosi1987exact,diosi2000robustness,diosi2002exact} and others \cite{paz1993reduction,brodier2004symplectic,eisert2004exact} in the context of a frequently studied special case of QBM, and generalize them to the symplectic covariant formalism.  We compare them to the preferred states and timescales associated with the POVM form for the dynamics derived above.  We do not necessarily expect closed-form expressions for arbitrary $\Fmat$ and $\Dmat$, but we can nevertheless show that the preferred quantities are well-defined and generally distinct.

The frequently studied special case of the dynamics can be described as momentum diffusion and spatial decoherence that is frictionless ($\gamma = 0$) and spatially homogeneous.  This is often called simply quantum Brownian motion, but we will call it \emph{pure spatial decoherence} to distinguish it from the general case, \eqref{eq:QBM-general}.  It is defined by setting $\Fmatin^{x}_{\phantom{x}p}=1/m$, $\Dmatin^{pp} = \DdiffDK$, and all other coefficients of $\Fmatin^{a}_{\phantom{a}b}$ and $\Dmatin^{ab}$ to zero. The dynamical equation for the Wigner function reduces to
\begin{align}
\label{eq:wig-decoh}
%\frac{\partial}{\partial t} W_t(x,p) = \left[-\frac{p}{ m} \frac{\partial}{\partial x} +  \frac{\DdiffDK}{2} \frac{\partial^2}{\partial p^2} \right] W_t(x,p),
\partial_t W_t(x,p) = \left[-\frac{p}{ m} \partial_x + \frac{\DdiffDK}{2} \partial^2_p \right] W_t(x,p).
\end{align}
Pure spatial decoherence is often obtained mathematically from an explicit model of the environment as a thermal bath of oscillators coupled linearly in $\hat{x}$, followed by taking the large-temperature limit \cite{unruh1989reduction,strunz2002decoherence,eisert2004exact}. It also well describes the dynamics taken by a test mass subjected to collisional decoherence \cite{joos1985emergence, gallis1990environmental, schlosshauer2008decoherence,diosi2002exact} from an environment of lighter particles \cite{hornberger2006master}, blackbody radiation \cite{joos1985emergence}, or low-mass dark matter \cite{riedel2013direct}.

In order to extend the results of Di\'{o}si and Kiefer to symplectic generality, we recall that the Husimi $Q$ function and (when it is well-defined) the Glauber $P$ function can, for any state $\ket{\psi}$, be usefully generalized \cite{klauder2007generalized} to
\begin{gather}
Q^{\ket{\psi}}_\rho (\alvec) \equiv \frac{1}{2 \pi} \matrixelement{\psi,\alvec}{ \rho }{\psi,\alvec},\\
\rho \equiv \int \! \dd{\alvec}\, P^{\ket{\psi}}_\rho(\alvec) \, \projector{\psi,\alvec},
\end{gather} 
where $\ket{\psi,\alvec} = \DisOp_{\alvec} \ket{\psi}$ is the state $\ket{\psi}$ translated in phase space by $\alvec$.  These reduce to the original $Q$ and $P$ functions when $\ket{\psi}$ is the coherent state centered at the origin in phase space, i.e., the ground state of the harmonic oscillator, $\ket{\IdT;\alvec=\mathbf{0}}$.  In the context of quadratic Hamiltonians, it is natural to concentrate on the case of generalized $Q$ and $P$ functions for which $\ket{\psi} = \ket{\CXTmat;\alvec=\mathbf{0}}$, i.e., a general Gaussian state centered at the origin with covariance matrix $\CXTmat$, and adopt the shorthand notation $Q^{\CXTmat}_\rho \equiv Q^{\ket{\CXTmat;\alvec=\mathbf{0}}}_\rho$ and $P^{\CXTmat}_\rho \equiv P^{\ket{\CXTmat;\alvec=\mathbf{0}}}_\rho$.  These can be related (see Appendix~\ref{sec:afour}) to the Wigner function by 
\begin{gather}
\label{eq:q-for-w}
Q^{\CXTmat}_{\rho} =  \Gfunc_{\CXTmat} \ast W_{\rho} ,\\
\label{eq:w-for-p}
W_{\rho} =  \Gfunc_{\CXTmat} \ast  P^{\CXTmat}_{\rho}
\end{gather} 

We can obtain the dynamical equations for  $Q_\rho^{\CXTmat}$ by making the replacement $W_\rho \to Q_\rho^{\CXTmat}$ and $\Dmat \to \Dmat^Q_{\CXTmat}$ in \eqref{eq:symplectic-wig-decoh}, where
\begin{align}
\Dmat^Q_{\CXTmat} \equiv \Dmat - (\Fmat \CXTmat + \CXTmat \Fmat^\intercal).
\end{align}
Likewise is true for $P_\rho^{\CXTmat}$, by making the replacement $W_\rho \to P_\rho^{\CXTmat}$ and $\Dmat \to \Dmat^P_{\CXTmat}$ in \eqref{eq:symplectic-wig-decoh}, where
\begin{align}
\Dmat^P_{\CXTmat} \equiv \Dmat + ( \Fmat \CXTmat + \CXTmat \Fmat^\intercal).
\end{align}
The solutions are
\begin{align}
\label{QBM-solution-Q}
Q^{\CXTmat}_{e^{\mathcal{L}t}[\rho]} =  \Gfunc_{\Cmat_t-\Emat^{\CXTmat}_t} \ast \left(e^{ 2 \gamma t}  Q^{\CXTmat}_{\rho} \circ e^{-t \Fmat} \right),\\
\label{QBM-solution-P}
P^{\CXTmat}_{e^{\mathcal{L}t}[\rho]} =  \Gfunc_{\Cmat_t+\Emat^{\CXTmat}_t} \ast \left(e^{ 2 \gamma t}  P^{\CXTmat}_{\rho} \circ e^{-t \Fmat} \right) 
\end{align}
where
\begin{align}
\begin{split}
\Emat^{\CXTmat}_t & \equiv e^{t\Fmat} \CXTmat e^{t\Fmat^\intercal} - \CXTmat \\
&= \int_0^t\! e^{t\Fmat} \left[\Fmat \CXTmat + \CXTmat \Fmat^\intercal\right] e^{t\Fmat^\intercal}  \dd{\tau}. 
\end{split}
\end{align}
These can be checked using (\ref{eq:q-for-w}--\ref{eq:w-for-p}), (\ref{eq:id1}--\ref{eq:id3}), and \eqref{QBM-solution}.

%%%%%%%%%%%%%%%%%%%%%%
\subsection{Unraveling the Glauber $P$ function}
%%%%%%%%%%%%%%%%%%%%%%

%%%     Figure:  range-of-alpha     %%%%%%%%%%%%%%%%%%%%%%%%%%%%%
\newcommand{\dookfactor}{0.90} %single column width adjustment

\begin{figure} [b]
	\centering
	\includegraphics[width=\dookfactor\columnwidth]{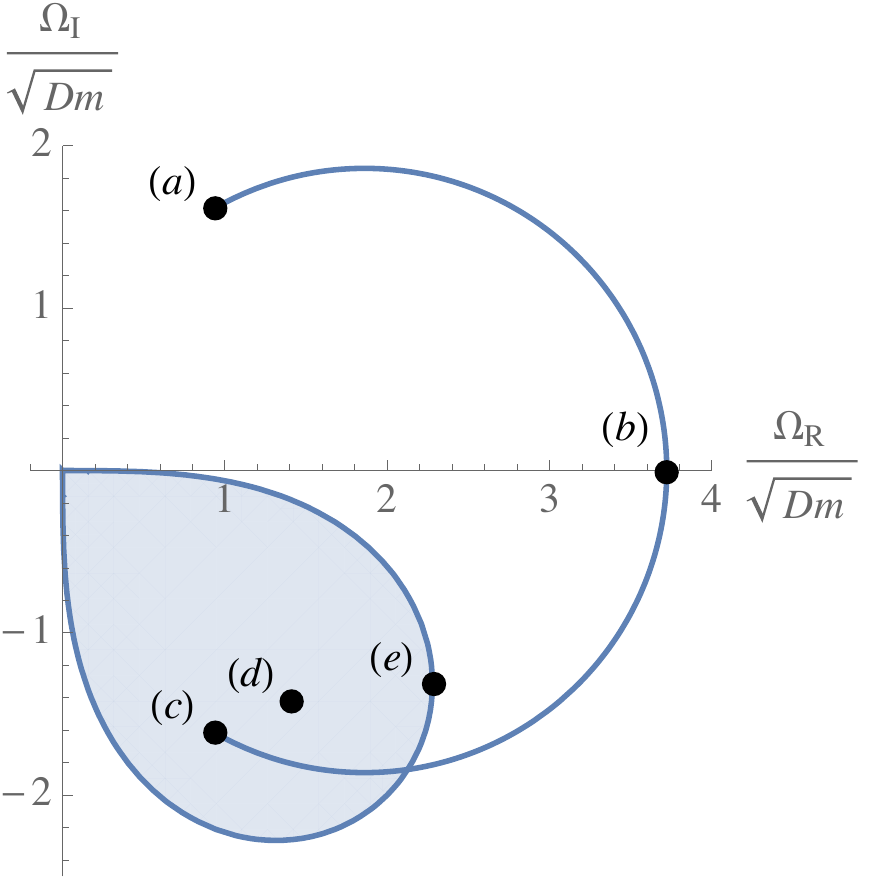}
	\caption{(Color online) Some candidate Gaussian pointer states with spatial wavefunction $\braket{x}{\CXTmat;\alvec = \mathbf{0}}  \propto \exp[-(\dar + i\dai) x^2/4]$ and corresponding covariance matrix given by \eqref{eq:CX-param}.  Wavefunction normalizability requires that $\dar > 0$.  The arc corresponds to the states $\CXTmat = \CVTmat_r$ compatible with a POVM form, \eqref{eq:POVM-general}, for the dynamics, \eqref{eq:wig-decoh}.  It is parametrized by $\rr \in [0,1]$, running down from (a) $\CVTmat_{1} = \CVTmat$ to (b) $\CVTmat_{1/2}$ to (c) $\CVTmat_{0} = \overline{\Rmat^{-1} \CVmat}$.  Under infinitesimal evolution, the state that can be best approximated by a pure-state according to the Hilbert-Schmidt norm is labeled by (d). States within the shaded region satisfy $\Dmat^P_{\CXTmat} \ge 0$, allowing for an interpretation in terms of a diffusing probability distribution $P^{\CXTmat}_\rho$ over pointer states.  The state producing the minimal linear entropy (equivalently, maximum $\dar$) from that region is labeled by (e).}
	\label{fig:range-of-alpha}
\end{figure}
%%%%%%%%%%%%%%%%%%%%%%%%%%%%%

As observed in Refs.\ \cite{diosi1987exact,diosi2000robustness}, \eqref{QBM-solution-P} is noteworthy because, when $P^{\CXTmat}_{e^{t\mathcal{L}}[\rho]} (\alvec)$ is defined, the system is described by a (classical) probability distribution diffusing over a set of pure Gaussian states with a preferred covariance matrix $\CXTmat$.   However, this interpretation is only viable when the diffusion matrix $\Dmat^P_{\CXTmat} \equiv \Dmat + ( \Fmat \CXTmat + \CXTmat \Fmat^\intercal)$ for the Glauber function $P_\rho^{\CXTmat}$ is positive semidefinite. This restriction $\Dmat^P_{\CXTmat} \ge 0$ defines a region in the space of possible Gaussian pure-state covariance matrices $\CXTmat$.  

Note that this region may be empty for some choices of dynamical parameters.\footnote{We observed numerical and analytic evidence that the region $\Dmat^P_{\CXTmat} > 0$ has strictly positive volume for all but a measure zero subset of the dynamical parameter space. That is, it appears that there is always a choice of $\CXTmat$ that allows for the diffusive interpretation with $\Dmat^P_{\CXTmat}$ for almost any dynamical parameters. However we could not find a proof of this.}  For instance, if $\Fmat^{x}_{\phantom{x}x} = \mu = - \Fmat^{p}_{\phantom{p}p}$ for real $\mu$ and all other parameters vanish, then no choice of  $\CXTmat$ makes $\Dmat^P_{\CXTmat}$ positive.  In this case, the dynamics continuously squeeze phase space toward one axis, so that an initial Gaussian state becomes arbitrarily squeezed with increasing time, and hence eventually not expressible as a mixture of Gaussians with fixed, finite covariance matrix $\CXTmat$.

If there is a region of compatible $\CXTmat$ with finite volume, it is then possible to look for an additional criterion that would prefer some states in this region over others.  One choice is to find the pure initial state $\ket{\psi}$ that minimizes the instantaneous linear entropy production $\dd S_{\mathrm{L}}/\dd t$, where $S_{\mathrm{L}} = 1-\Tr\rho^2$.  This is motivated by the intuitive notion of the \emph{predictability sieve}; pointer states are the quantum states that are most stable under interactions with the environment \cite{zurek1993preferred,paz1993reduction,dalvit2005predictability}, and thereby produce little entanglement entropy.

In general, the linear entropy production is minimized by choosing the most squeezed dimension of the initial state to be along the largest eigenvector of the diffusion matrix $\Dmat$.  We follow Kiefer et al \cite{kiefer2007pointer} and Di\'{o}si and Kiefer \cite{diosi2000robustness} and parametrize the most general unit-determinant, positive semidefinite matrix $\CXTmat$ according to the complex number $\da = \dar + i \dai$ (with $\dar > 0$):
\begin{align}
\label{eq:CX-param}
\CXTmat = \frac{1}{\dar}\left( \begin{array}{cc} 1 & - \dai/2 \\ - \dai/2 & \abs{\da}^2/4 \end{array}\right)
\end{align}
This corresponds to a spatial wavefunction $\braket{x}{\CXTmat;\alvec = \mathbf{0}} = (\dar/2\pi)^{1/4} \exp(-\da x^2/4)$. In the special case of pure spatial decoherence, the linear entropy production is proportional to $\dar$, and one can check for pure spatial decoherence \eqref{eq:wig-decoh} that $\dar$ is maximized by the choice $\da = 3^{1/4} (3^{1/2} - i) \sqrt{\DdiffDK m}$ under the constraint that $\Dmat^P_{\CXTmat}$ is positive \cite{diosi2000robustness}.

One may alternatively consider the pointer state selected by the principle of \emph{Hilbert-Schmidt robustness} \cite{diosi1987exact,diosi1988quantum,gisin1995relevant,diosi2000robustness,busse2010pointer, soergel2015unraveling}.  The time-dependent pure states $\ket{\psi_t}$ that best approximate, according to the Hilbert-Schmidt norm, the impure state $\rho_t$ evolving under QBM can be shown to solve \cite{diosi1987exact}
\begin{align}
\label{eq:stupid}
\frac{\dd}{\dd t}\ket{\psi_t} = \Big( \mathcal{L} \big[\projector{\psi_t}\big] - \matrixelement{\psi_t}{\mathcal{L} \big[\projector{\psi_t}\big]}{\psi_t} \Big) \ket{\psi_t}.
\end{align}
For pure spatial decoherence \eqref{eq:wig-decoh}, the unique stationary solution to this non-linear equation are candidate pointer states \cite{diosi1986stochastic,diosi1987exact,gisin1995relevant,diosi2000robustness,busse2009emergence,busse2010pointer}.  They are all equivalent up to translations in phase space, being given by a Gaussian wavepacket with a covariance matrix specified by $\da = (1 - i) \sqrt{2 \DdiffDK m}$ \cite{diosi2000robustness}.

In Fig.\ \ref{fig:range-of-alpha}, these two preferred states and the preferred region associated with the condition $\Dmat^P_{\CXTmat} \ge 0$ are compared with the one-parameter family of Gaussian states characterized by the covariance matrix $\CVTmat_\rr$ given by \eqref{eq:POVM-general-mat} that are compatible with the general POVM form \eqref{eq:POVM-general} for pure spatial decoherence.

%%%%%%%%%%%%%%%%%%%%%%
\subsection{Positivity times}
%%%%%%%%%%%%%%%%%%%%%%

Di\'{o}si and Kiefer also calculated \cite{diosi2002exact} the characteristic times $T_W$ and $T_P$ at which the Wigner function $W_\rho$ and the traditional Glauber function $P_\rho = P^{\IdT}_\rho$ of an arbitrary quantum state became strictly positive under pure spatial decoherence \eqref{eq:wig-decoh}.  As they conjectured was possible, this was extended to symplectically general QBM dynamics by Brodier and Ozorio de Aleida \cite{brodier2004symplectic}. The Wigner positivity time $T_W$ is strictly a property of the dynamics in the sense that the time is independent of the initial Wigner function (so long as it is pure and not already positive) \cite{brodier2004symplectic}.  Once a Wigner function is positive, it remains so indefinitely under QBM dynamics.  Here we collect these results and likewise treat the positivity of the Glauber $P_{\rho}^{\CXTmat}$ function, corresponding to Gaussian kernels with arbitrary covariance matrix $\CXTmat$, in symplectic generality.

First note that
\begin{align}
\begin{split}
\label{eq:rejigger-wig}
\Wig{e^{\mathcal{L}T}[\rho]} &=  \Gfunc_{\Cmat_t} \ast \left(e^{ 2 \gamma t}  \Wig{\rho} \circ e^{-t \Fmat} \right) \\
&=  \left( ( \Gfunc_{\Cmat_t} \circ e^{t \Fmat} ) \ast  \Wig{\rho} \right) \circ e^{-t \Fmat}\\
&=   e^{ 2 \gamma t} \left(  \Gfunc_{e^{-t \Fmat} \Cmat_t e^{-t \Fmat^\intercal}}  \ast  \Wig{\rho} \right) \circ e^{-t \Fmat}.
\end{split}
\end{align}
The key idea is that convolving the Wigner function by a Gaussian $\Gfunc_{\IdT}$ yields the Husimi $Q$ function, and the latter is always positive \cite{diosi2002exact}, but convolving with a sharper Gaussian (e.g., $\Gfunc_{\Id/4}$) will never produce a positive function from a nonpositive pure state (or vice versa) \cite{brodier2004symplectic}.
The classical flow $e^{-t \Fmat}$ and the multiplicative factor $e^{ 2 \gamma t} >0$ do not change the positivity of a Wigner function, so convolution by $\Gfunc_{\Cmat_t}$ will make a Wigner function for a pure state positive if and only if $\jdet{\Cmat_t} \ge \jdet{\IdT} =  1/4$.

%%%     Figure:  Strobe POVM     %%%%%%%%%%%%%%%%%%%%%%%%%%%%%
\newcommand{\introdfactor}{0.8} %single column width adjustment

\begin{figure*} [bt]
	\centering
	\includegraphics[width=\introdfactor\textwidth]{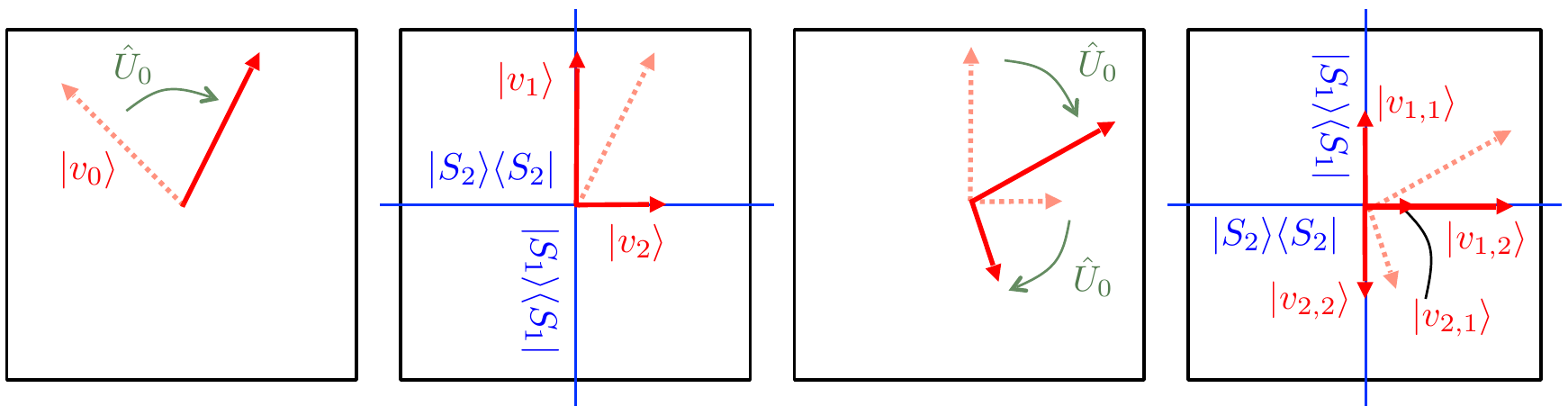}
	\caption{(Color online) The real part of the Hilbert space of a two-dimensional system.  The system evolves stroboscopically according to a unitary $\hat{U}_0$ which is periodically interrupted every time step $T$ by a projective measurement in the basis $\{ \ket{S_1}, \ket{S_2}\}$.  After two measurements, there are four possible outcomes: $(1,1)$, $(1,2)$, $(2,1)$, and $(2,2)$, corresponding to the final states $\ket{v_{ij}} = \projector{S_j}\hat{U}_0\projector{S_i}\hat{U}_0\ket{v_0}$.  The basis $\{\ket{S_i}\}$ plays the role of a pointer basis, but this evolution is equivalent to periodic measurements in the basis $\{{\hat{U}}_0^{-(T-t)/T} \ket{S_i}\}$ with a constant time offset $t \in [0,T]$ (whether or not the outcomes $i$ are remembered or forgotten).  In this sense, the pointer basis for this system is only sensibly specified relative to an offset time, i.e., as a pair $(t,\{\hat{U}_0^{-(T-t)/T} \ket{S_i}\})$.}
	\label{fig:gauge}
\end{figure*}
%%%%%%%%%%%%%%%%%%%%%%%%%%%%%

Thus by examining \eqref{eq:rejigger-wig}, $T_W$ is defined by
\begin{align}
\jdet{\Cmat_{T_W}} = \frac{1}{4}e^{-4 \gamma T_W}.
\end{align}
Using \eqref{eq:det-c-bound}, once can show that $T_W$ is finite (except for the extremal case $\gamma = - \sqrt{\jdet \Dmat }$).
Likewise, the generalized Glauber $P$ function associated with a set of preferred Gaussian state with covariance matrix $\CXTmat$ is given (when it exists) by $W_\rho = \Gfunc_{\CXTmat} \ast P^{\CXTmat}_{\rho}$, so
\begin{align}
\begin{split}
P^{\CXTmat}_{e^{\mathcal{L}T}[\rho]} &=  \Gfunc_{\Cmat_t - \CXTmat} \ast \left(e^{ 2 \gamma t}  \Wig{\rho} \circ e^{-t \Fmat} \right) \\
&=   e^{ 2 \gamma t} \left(  \Gfunc_{e^{-t \Fmat} (\Cmat_t -\CXTmat) e^{-t \Fmat^\intercal}}  \ast  \Wig{\rho} \right) \circ e^{-t \Fmat}.
\end{split}
\end{align}
It is guaranteed to exist and be positive at the characteristic time $T_P^{\CXTmat}$ defined by 
\begin{align}
\jdet{\Cmat_{T^{\CXTmat}_P} - \CXTmat} = \frac{1}{4} e^{-4 \gamma T^{\CXTmat}_P}.
\end{align}
Different choices of the preferred covariance matrix $\CXTmat$ lead to different times upon which the associated Glauber function $P^{\CXTmat}_\rho$ becomes positive.  For pure spatial decoherence \eqref{eq:wig-decoh}, $T_W = {3}^{1/4} \sqrt{m/\DdiffDK} = T/\sqrt{2}$ \cite{diosi2002exact}, and $T^{\CXTmat}_P$ is the solution to $12 m^2 + 6 m \dai t + (\dai^2  + \dar^2) t^2 - \DdiffDK \dar t^3=0$.

Many other extensions are possible.  For instance, it is clear that one could calculate the time at which the Cahill $R$ function, which continuously interpolates between the $P$, $W$, and $Q$ function \cite{lee1991measure}, becomes positive for different values of the interpolation parameter.  Likewise one could define pointer states of the system to be the Gaussians with covariance matrix $\CXTmat$ such that diffusion in the Glauber function $P^{\CXTmat}_{\rho}$ is preferred according to a criterion other than minimizing the linear entropy production.  None of these stand out as definitive notions of classicality.

It is worth emphasizing that the condition $\Wig{\rho} \ge 0$ is a property strictly of the quantum state $\rho$ itself, whereas the condition $P^{\CXTmat}_{\rho} \ge 0$ (and most other criteria based on the functions $P^{\CXTmat}_{\rho}$ or $Q^{\CXTmat}_{\rho}$) are dependent on the \emph{dimensionful} choice $\CXTmat$.  The traditional definitions for the Glauber and Husimi functions $P_{\rho}=P^{\IdT}_{\rho}$ and $Q_{\rho} = Q^{\IdT}_{\rho}$ do not avoid this because they depend implicitly on a length scale $\sigma_x$ used to define the identity matrix, 
\begin{align}
%\frac{1}{2} \Id = 
\IdT = \frac{1}{2} \left( \begin{array}{cc} 1 & 0 \\ 0 & 1 \end{array}\right) = \left( \begin{array}{cc} \sigma_x^2 & 0 \\ 0 & \sigma_p^2 \end{array}\right)  = \left( \begin{array}{cc} \sigma_x^2 & 0 \\ 0 & \hbar^2/4\sigma_x^2 \end{array}\right)
\end{align}
where $\sigma_x^2 = \avg{\hat{x}^2} - \avg{\hat{x}}^2= \matrixelement{\alvec=0}{\hat{x}^2}{\alvec=0}$ is the spatial variance of the coherent states.  This length scale is usually taken from the dynamics (most often, the width of the oscillator potential) and is not a property of the state $\rho$ alone.  Likewise, it is often taken for granted that squeezed states are nonclassical (e.g., \cite{diosi2002exact}), but squeezing is always relative to an assumed scale separate from $\rho$.

%%%%%%%%%%%%%%%%%%%%%%%%%%%%%
\section{Discussion}
%%%%%%%%%%%%%%%%%%%%%%%%%%%%%

%%%%%%%%%%%%%%%%%%%%%%%%%%%%%%
%\subsection{Rotating frame}
%%%%%%%%%%%%%%%%%%%%%%%%%%%%%%

General QBM dynamics arise from the lowest order terms in the Taylor series expansion of a smooth Hamiltonian for a Markovian open system, giving them similar conceptual importance and pedagogical usefulness as the harmonic oscillator has in the study of closed quantum system. The linearity of the Lindblad operators means their influence can be described as continuous weak monitoring of the phase-space variable $\hat{\alpha} = (\hat{x},\hat{p})$ \cite{barnett2005quantum,chia2011complete,wiseman2014quantum}.  In this work we have shown how these weak measurements add up to a single strong measurement, on a timescale that characterizes the dynamics, in the form of a POVM-and-prepare (entanglement-breaking) quantum channel.  The symplectic generality exhibited here will be very important for extending, to Markovian open systems, existing quasiclassicality theorems \cite{hepp1974classical,hagedorn1981semiclassical,hagedorn1985semiclassical,combescure2012coherent} that apply to evolution generated by \emph{any} closed-system Hamiltonian that is sufficiently smooth to be treated as approximately locally quadratic, rather than just for representative toy environments like baths of harmonic oscillators. 

The various forms exhibited in \eqref{eq:POVM-general} suggest that the overcomplete set of Gaussian states forming the POVM are only defined up to a sort of gauge freedom
\begin{align}
\label{eq:POVM-freedom}
%\{\hat{U}_{e^{\Hmat t}}^\dagger
\Big(0,\Big\{\ket{\CVTmat;\alvec}\Big\}\Big) \to \Big(t,\Big\{\ket{\overline{\Rmat^{- t}\CVmat}; \alvec}\Big\}\Big)
\end{align}
for any $t \in [0,T]$.  On the other hand, the time scale $T$ associated with the dynamics is independent of this freedom \footnote{Of course, \emph{all} POVMs with respect to the frame of Gaussian states with a covariance matrix $\CXTmat$ are equivalent if we allow them to be supplemented with an \emph{arbitrary} unitary immediately before and after the measurement.  The one-parameter family in \eqref{eq:POVM-freedom} is notable because it is merely inserted at some point in the normal unitary component of the evolution.}.

This is the continuum analog to a preferred basis ambiguity that can be found in the simpler case where the dynamics are described stroboscopically by unitary evolution punctuated by a simple projective measurement of a discrete variable. See Fig.\ \ref{fig:gauge}.  The ambiguity arises because of the periodic measurement events and the non-trivial unitary evolution of the system in between them; it is not a special property of continuous system.  

Given this freedom, as well as the alternative pointer state criteria discussed in the previous section, it is not clear whether the pointer states of quantum Brownian motion are best understood in terms of the POVM form for the dynamics presented above.  However, this basis ambiguity may play a conceptual role in any future satisfactory notion of pointer states when decoherence is taking place alongside unitary evolution.  In other words, one should be suspicious of the intuition that there is a single true preferred basis (whether overcomplete or otherwise) that one might develop from studying simple models of pure decoherence in an orthonormal basis.
 
Although the characteristic time $T$ of the POVM description differs (by a factor of order unity) from the exact times $T_W$ and $T_P$ at which the Wigner and Glauber functions become positive, the positivity manifestly implied by the former description is arguably more transparent; the Wigner function following a POVM-and-prepare channel in a Gaussian state basis is obviously positive.  It also suggests new approaches to understanding phase-space positivity, or the quantum-classical description more generally, in continuous-variable systems that aren't described by QBM.
 
The symplectic covariance of our results offers illuminating generality compared to earlier discussion of pointer states in special cases of QBM \cite{dalvit2005predictability,diosi2000robustness,unruh1989reduction,zurek1993coherent,strunz2002decoherence}.  
For example, the pointer states associated with the POVM form for the damped harmonic oscillator dynamics are a generalization, to arbitrary damping, of the coherent states that were identified as pointer states in the underdamped limit ($\gamma \ll \omega$) \cite{dalvit2005predictability}.

It is notable that, except in the extremal case $\gamma^2 = \abs{\Dmat}$, the preferred states associated with the POVM form for the QBM dynamics are always well defined and unambiguous, a result which may also apply to pointer states associated with Hilbert-Schmidt robustness.  In contrast, the predictability sieve often produces singular pointer states like the position eigenstates with divergent momentum dispersion \cite{dalvit2005predictability}, unless supplemented with additional cumbersome principles such as a Glauber $P$ function dispersion interpretation, or a finite-time averaging scheme.  Of course, no elegant principle exists that unambiguously identifies sensible pointer states (or their nonexistence) for arbitrary dynamics, and the predictability sieve appears to offer more guidance there.

Our most restrictive assumption has been that the dynamics are Markovian and time homogeneous.  One way to relax this is by allowing $\Dmat$ and $\Fmat$ to vary with time, possibly in a way that depends on the initial state.  This will lead to straightforward modifications of $\Cmat_t$, $T$, and $\CVTmat$, and it is still possible to describe the dynamics stroboscopically as a Gaussian-state POVM-and-prepare channel \cite{ShuyiWriteup}.  In this case the preferred states and timescales are generally not determined solely by the dynamics, but also by the initial state.  

This might be extended to cover more general models of non-Markovian dynamics, like the finite-temperature bath of linearly coupled oscillators of Caldeira-Leggett \cite{caldeira1983path, unruh1989reduction, hu1992quantum}.  However, in such cases $\Dmat$ is not necessarily positive-definite, and this can interfere with constructing the POVM form.  Indeed, this form must breakdown when the memory of the environment exceeds the stroboscopic time interval.  (Trivially, a finite bath has a finite global recurrence time and so will eventually restore any initial non-classical superposition states of the system.)

It would be especially interesting to see if pointer states can be identified in non-Markovian dynamics such that the environment's memory consists only of the classical history of those preferred state, e.g., if the evolution can be described as an iterated sequence of POVM-and-prepare channels depending on previous outcomes.  On the other hand, it is hard to see how pointer states could be usefully defined for the more general non-Markovian case where the coherence information \emph{between} pointer states feeds back from the environment into the system; in that case, one would rather say that decoherence had not been effective and there simply are no pointer states.

%%%%%%%%%%%%%%%%%%%%%%%%%%%%%
%       Bibliography        %
%%%%%%%%%%%%%%%%%%%%%%%%%%%%%
\bibliographystyle{apsrev4-1}
\bibliography{zotriedel,mylocalbib}
%%%%%%%%%%%%%%%%%%%%%%%%%%%%%

%%%%%%%%%%%%%%%%%%%%%%%%%%%%%%
\section{Acknowledgements}
%%%%%%%%%%%%%%%%%%%%%%%%%%%%%%

I thank Shuyi Zhang for pointing out the connection to Weyl spinors, Joshua Combes for suggesting the example in Appendix B, and Adolfo del Campo, Lajos Di\'{o}si, Claus Kiefer, Gordan Krnjaic, Wojciech Zurek, and Michael Zwolak for discussion.  Research at the Perimeter Institute is supported by the Government of Canada through Industry Canada and by the Province of Ontario through the Ministry of Research and Innovation.  This work was also supported by the John Templeton Foundation through Grant No. 21484.

%%%%%%%%%%%%%%%%%%%%%%%%%%%%%%
\appendix
\section{Symplectic QBM}
%%%%%%%%%%%%%%%%%%%%%%%%%%%%%%

In this appendix we briefly review QBM along the lines of Brodier and Ozorio de Almeida \cite{brodier2004symplectic} and Robert \cite{robert2012time}, but in context and notation better suited for our purposes that especially emphasizes manifest symplectic covariance.  See also the appendix to Ref.~\cite{riedel2015decoherence}, Chapter 6 of Ref. \cite{wiseman2014quantum}, and a forthcoming pedagogical treatment \cite{ShuyiWriteup}.

%%%%%%%%%%%%%%%%%%%%%%%%%%%%%%%%%%%%%%%%
\subsection{Linear symplectic transformations}
%%%%%%%%%%%%%%%%%%%%%%%%%%%%%%%%%%%%%%%%

For a single continuous classical degree of freedom, phase space is a two-dimensional vector space equipped with the symplectic form as represented by the antisymmetric Levi-Civita symbol $\epsilon^{ab}$ (with $\epsilon^{xp} = \epsilon_{px} = +1 = - \epsilon^{px} = - \epsilon_{xp}$).  Analogously to Lorentz indices, symplectic indices are raised and lowered by contracting with the second index in the symplectic form.  Unlike for a Lorentzian metric, which is symmetric, the anti-symmetry of the Levi-Civita symbol means there is an overall sign flip depending on which index is raised and which is lowered in a contraction: $X_{a} Y^a = \epsilon_{ab} X^b Y^a = -\epsilon_{ba} X^b Y^a = -X^b Y_b$.  The symplectic indices behave just like Weyl spinor indices, which are reviewed in many introductory quantum field theory textbooks (e.g., Ref~\cite{srednicki2007quantum}).

Classical Hamiltonian evolution is given by a time-parametrized family of symplectomorphisms (canonical transformations) on phase space, which are characterized by the fact that the Jacobian of the transformation preserves the symplectic form at each point.  We concentrate on the local dynamics of smooth Hamiltonians, so we are most interested in the symplectic \emph{linear} transformations, which, for one degree of freedom, are effected with the Lie group of 2-by-2 real matrices $\CXmatin^a_{\phantom{a}b}$ with unit determinant, $\mathrm{SL}_2(\mathbb{R})$.  These preserve the symplectic form: $\epsilon^{ab} = \CXmatin^a_{\phantom{a}c} \epsilon^{cd} \CXmatin^b_{\phantom{b}d}$ or, equivalently, $\boldsymbol{\epsilon} = \CXmat \boldsymbol{\epsilon}  \CXmat^\intercal$.

Each (classical) symplectic linear transformation $\CXmat$ is associated with a corresponding quantum unitary transformation  $\hat{U}_{\CXmat} = \exp(-i \lnVmatin_{ab} \hat{\alpha}^a\hat{\alpha}^b/2)$, where $\lnVmat$ is the matrix $\lnVmatin^a_{\phantom{a}b} = \epsilon^{ac} \lnVmatin_{cb}$ and $\CXmat = e^{\lnVmat}$.  In particular, $\avg{\hat{U}_{\CXmat}^\dagger \hat{\alvec}\hat{U}_{\CXmat}} = \CXmat \avg{\hat{\alvec}}$ and $\hat{U}_{\CXmat^{-1}} = \hat{U}_{\CXmat}^{-1} = \hat{U}_{\CXmat}^{\dagger}$. When this unitary acts on a coherent states $\ket{\alvec} = \ket{\Id/2; \alvec}$, it transforms it \cite{weedbrook2012gaussian, combescure2012coherent} into a (generally squeezed) Gaussian state centered on the classically shifted point in phase space: $\hat{U}_\CXmat \ket{\alvec}= \ket{\CXTmat;\CXmat \alvec}$, where $\CXTmat \equiv \CXmat \CXmat^\intercal/2$.  Here, the pure Gaussian states $\ket{\CXTmat;\alvec}$  are parametrized by their mean $\alvec \equiv \avg{\hat{\alvec}}$ and their 2-by-2 positive definite, unit-determinant covariance matrix $\CXTmat$ with elements $\CXTmatin^{ab} = \avg{\hat{\alpha}^a \hat{\alpha}^b + \hat{\alpha}^b \hat{\alpha}^a}/2$.  ($\CXTmat$ determines $\ket{\CXTmat;\alvec}$ up to a phase, which is sufficient for our purposes.)  The many-to-one mapping $\CXmat \to \CXTmat = \CXmat \CXmat^\intercal/2$ collapses the three dimensional Lie group $\mathrm{SL}_2(\mathbb{R})$ down to a two-dimensional manifold that can be parametrized as in \eqref{eq:CX-param}.

(Note that in the context of Gaussian quantum phase-space distributions, some authors differ from our convention by setting $\hbar = 2$.  In this case, the coherent state satisfies $\langle \hat{x}^2 \rangle \langle \hat{p}^2 \rangle = \sigma^2_x \sigma^2_p = \hbar^2/4 = 1$, i.e., the covariance matrix is $\Id$ rather than $\IdT=\Id/2$ \cite{weedbrook2012gaussian}.)

%%%%%%%%%%%%%%%%%%%%%%%%%%%%%%%%%%%%%%%%
\subsection{Quantum Brownian motion}
%%%%%%%%%%%%%%%%%%%%%%%%%%%%%%%%%%%%%%%%

A single quantum continuous degree of freedom undergoing open-system, time-homogeneous, and Markovian dynamics, forms a quantum dynamical semigroup \cite{lindblad1976generators, alicki2007quantum} described by a Lindblad master equation
\begin{align}
\label{eq:lindblad}
\partial_t \rho_t = - i [\hat{H},\rho_t] +\sum_i \left[ \hat{L}^{(i)} \rho_t \hat{L}^{(i) \dagger} - \frac{1}{2} \{ \hat{L}^{(i) \dagger} \hat{L}^{(i)} , \rho_t \} \right].
\end{align}
Ideal QBM is the special case when the Hamiltonian is quadratic and the Lindblad operators are linear with the phase-space operators: $\hat{H} =\frac{1}{2}  \Hmatin_{ab} \hat{\alpha}^a \hat{\alpha}^b$ and $\hat{L}^{(i)} = L^{(i)}_{a} \hat{\alpha}^a$, where $\Hmatin_{ab}$ is a real symmetric matrix and the $L^{(i)}_{a}$ are complex vectors.  (Linear terms in the Hamiltonian either can be handed explicitly separately \cite{robert2012time} or, so long as the quadratic terms $\Hmatin_{ab}$ are nonzero, can be eliminate through a phase-space translation $(x,p) \to (x+x_0,p+p_0)$.) We can then change variables to $\Dmatin_{ab} =  \mathrm{Re} \sum_i  (L^{(i)}_{a})^* L^{(i)}_{b}$, $\gamma =   \frac{1}{2i} \epsilon^{ab} \sum_i  (L^{(i)}_{a})^* L^{(i)}_{b} = \mathrm{Im} \sum_i (L^{(i)}_{x})^* L^{(i)}_{p} $, and $\Fmatin_{ab} = \Hmatin_{ab} + \epsilon_{ab} \gamma$. (Note that our convention for the matrix $\Dmatin_{ab}$ agrees with Di\'{o}si and Kiefer \cite{diosi2000robustness, diosi2002exact}, but differs by a factor of two from Isar et al., \cite{isar1994open}, Dekker and Valsakumar \cite{dekker1984fundamental}, and others.)

In the traditional Hilbert space representation this yields
\begin{align}
\label{eq:QBM-general-app}
\frac{\partial}{\partial t} \rho_t = -\frac{i}{2} \Fmatin_{ab}  \left[\hat{\alpha}^a,\left\{ \hat{\alpha}^b, \rho_t \right\} \right] - \frac{1}{2} \Dmatin_{ab} \left[ \hat{\alpha}^a,\left[\hat{\alpha}^b, \rho_t \right] \right],
\end{align}
Using the Wigner function
\begin{align}
%W_\rho (x,p) \equiv \frac{1}{2\pi} \int \dd{i\Delta x} e^{-p\Delta x} \matrixelement{x+\frac{\Delta x}{2}}{\rho}{x-\frac{\Delta x}{2}}
W_\rho (x,p) \equiv \frac{1}{2\pi} \int \dd{\Delta x} e^{-ip\Delta x} \matrixelement{x+\Delta x/2}{\rho}{x-\Delta x/2}
\end{align}
we can instead express these dynamics in the Wigner representation as a Fokker-Planck equation
\begin{align}
\label{eq:symplectic-wig-decoh-app}
\partial_t W_\rho (\alpha) = \left[ -\Fmatin^{a}_{\phantom{a}b} \partial_a \alpha^b + \frac{1}{2} \Dmatin^{ab} \partial_a \partial_b \right] W_\rho(\alpha) .
\end{align}
Because the equation for the \emph{classical} phase-space probability distribution under ideal Brownian motion is identical to \eqref{eq:symplectic-wig-decoh-app}, one can directly read off an interpretation of the coefficients.  In the absence of diffusion ($\Dmat = 0$), the classical equations of motion are $\dot{\alvec} = \Fmat \alvec$, with $e^{t \Fmat} = e^{-t\gamma}e^{t \Hmat}$ the classical (possibly dissipative) flow and $e^{t \Hmat}$ the Hamiltonian component.  The rate of dissipation is $\gamma$, where $\gamma<0$ implies that the environment is pumping energy into the system.  The (strictly quantum) constraint $\abs{\Dmat} \ge \hbar^2 \gamma^2$ ensures that diffusion is always sufficiently strong to prevent the dissapative contraction $e^{-t \gamma}$ from producing violations of the uncertainty principle.

%%%%%%%%%%%%%%%%%%%%%%%%%%%%%%%%%%%%%%%%
%\subsection{Special cases}
%%%%%%%%%%%%%%%%%%%%%%%%%%%%%%%%%%%%%%%%

We briefly mention the most important special cases.  (For others, see especially Ref.\ \cite{isar1994open}.) Normal friction (drag) on a free particle of the form $\ddot{x} = -(\lambda /m) \dot{x}$ is obtained with $\Fmatin^p_{\phantom{p}p} = -\lambda /m <0$, $\Fmatin^x_{\phantom{p}p} = 1/m >0$, and all other parameters zero. These dynamics are dissipative since $\gamma = -\Fmatin^a_{\phantom{a}a} = \lambda /m >0$.  A harmonic oscillator potential is represented by $\Fmatin^p_{\phantom{p}x} = -m \omega^2$, and the degree of damping (under- or over-damped) is controlled by $\lambda$.

Einstein-Smoluchowski diffusion is obtained when normal friction is supplemented with momentum diffusion ($\Dmatin^{pp} = 2\Ddiff$), since the motion is caused by many small momentum transfers from molecular collisions.  However, it takes place in the noninterial limit where the relaxation timescale $\gamma^{-1} = m/\lambda$ is short compared to the timescale on which observations are made. The fast relaxation means the momentum cannot grow, so on large scales the position acts as a random walk, $\avg{x^2} \propto T$. Contrast this with the frictionless momentum diffusion dynamics (pure spatial decoherence) described by \eqref{eq:wig-decoh}, for which the position variance $\avg{x^2}$ grows like $T^3$ rather than $T$.

%%%%%%%%%%%%%%%%%%%%%%%%%%%%%%%%%%%%%%%%
\subsection{Symplectic covariance}
%%%%%%%%%%%%%%%%%%%%%%%%%%%%%%%%%%%%%%%%

Under a linear symplectic transformation $\CXmatin^a_{\phantom{a}b}$, an arbitrary symplectic tensor $C^{ab\cdots}_{\phantom{ab\cdots}gh\cdots}$ with both covariant (lower) indices and contravariant (upper) indices transforms as
\begin{align}
C^{ab\cdots}_{\phantom{ab\cdots}gh\cdots} \overset{\CXmatin}{\longrightarrow} C^{a'b'\cdots}_{\phantom{a'b'\cdots}g'h'\cdots} (\CXmatin^{-1})^a_{\phantom{a}a'}(\CXmatin^{-1})^b_{\phantom{b}b'}\cdots \CXmatin^{g'}_{\phantom{g'}g}\CXmatin^{h'}_{\phantom{h'}h}\cdots
\end{align}
An equation with all upper and lower indices contracted together is automatically invariant under a (linear) symplectic transformation. For example, 
\begin{align}
\begin{split}
%A^c B_c \overset{\CXmatin}{\longrightarrow} A^{a} (\CXmatin^{-1})^c_{\phantom{c}a } \CXmatin^b_{\phantom{b}c } B_b = A^c B_c
A_c B^c \overset{\CXmatin}{\longrightarrow} A_{a}  \CXmatin^a_{\phantom{a}c } (\CXmatin^{-1})^c_{\phantom{c}b } B^b = A_c B^c
\end{split}
\end{align}  Dynamical equations \eqref{eq:QBM-general} and \eqref{eq:symplectic-wig-decoh-app} exhibit such manifest symplectic covariance.  

The operations of lowering and raising indices with $\epsilon^{ab}$ are compatible with an overall linear symplectic transformations because they are characterized by their preservation of the symplectic form, $\epsilon^{ab} = \CXmatin^a_{\phantom{a}c}\epsilon^{cd}\CXmatin^c_{\phantom{c}d}$.  (The Wigner function is a scalar phase-space density which transform trivially under a linear symplectic transformation.)

%%%%%%%%%%%%%%%%%%%%%%%%%%%%%%%%%%%%%%%%
\subsection{Generalized phase-space distributions}
%%%%%%%%%%%%%%%%%%%%%%%%%%%%%%%%%%%%%%%%
\label{sec:afour}

In the case of a preferred covariance matrix $\CXTmat$, the generalized Husimi $Q$ and Glauber $P$ functions are
\begin{gather}
	Q^{\CXTmat}_\rho (\alvec) \equiv \frac{1}{2 \pi} \matrixelement{\CXTmat;\alvec}{ \rho }{\CXTmat;\alvec},\\
	\rho \equiv \int \! \dd{\alvec}\, P^{\CXTmat}_\rho(\alvec) \, \projector{\CXTmat;\alvec}
\end{gather} 
These can be related to the Wigner function by first considering an arbitrary pure state $\rho = \projector{\chi}$ and calculating
\begin{align}
\begin{split}
Q^{\CXTmat}_{\projector{\chi}} (\alvec) &= \frac{1}{2 \pi} \abs{ \braket{ \CXTmat;\alvec }{\chi}}^2 \\
&= \frac{1}{2 \pi} \abs{ \matrixelement{\CXmat^{-1} \alvec}{\hat{U}_\CXmat^\dagger}{\chi}}^2\\
&= \int \! \dd{\betvec} W_{\projector{\CXmat^{-1} \alvec}}(\betvec) W_{\UCP_{\CXmat^{-1}}[\projector{\chi}]}(\betvec)\\
&= \int \! \dd{\betvec} \Gfunc_{\IdT}(\CXmat^{-1} \alvec - \betvec) W_{\projector{\chi}}(\CXmat \betvec)\\
&= \int \! \dd{\muvec} \Gfunc_{\IdT}(\CXmat^{-1} (\alvec - \muvec)) W_{\projector{\chi}}(\muvec)\\
&= \left( \Gfunc_{\CXTmat} \ast W_{\projector{\chi}} \right) (\alvec)
\end{split}
\end{align} 
where we have made use of the fact that $\Tr[\rho \rho'] = 2\pi\int \! \dd{\alvec} W_\rho(\alvec) W_{\rho'}(\alvec)$.  By linearity we can extend this to any mixed state $\rho = \sum_\psi p_\psi \projector{\psi}$:
\begin{align}
\label{eq:gen-Q-func}
Q^{\CXTmat}_{\rho} =  \Gfunc_{\CXTmat} \ast W_{\rho}.
\end{align} 
Likewise, when the generalized Glauber $P$ function exists it satisfies
\begin{align}
W_{\rho} =  \Gfunc_{\CXTmat} \ast  P^{\CXTmat}_{\rho}.
\end{align} 

%%%%%%%%%%%%%%%%%%%%%%%%%%%%%%
\section{Damped harmonic oscillator example}
%%%%%%%%%%%%%%%%%%%%%%%%%%%%%%

Here we calculate the characteristic time $T$ and the covariance matrix $\CVTmat$ describing the POVM-and-prepare channel associated with the common case of a lightly damped harmonic oscillator.  For a concrete example, consider the center-of-mass motion of the single cooled ion in a Paul trap described by Bushev et al. \cite{bushev2006feedback}.  In the absence of feedback, the master equation is 
\begin{align}
\label{eq:example-master}
\partial_t \rho &= -i \nu [{\hat{a}}^\dagger \hat{a}, \rho] + \Gamma (N+1) \mathcal{D}_{\hat{a}} [\rho] + \Gamma N \mathcal{D}_{{\hat{a}}^\dagger}[\rho]
\end{align} 
where $\hat{a} = (\hat{x}\sqrt{m \nu}+i\hat{p}/\sqrt{m \nu})/\sqrt{2}$ is the harmonic trap lowering operator, $m$ is the ion mass, $\nu = 1\, \mathrm{MHz}$ is the trap frequency, $\Gamma = 400\, \mathrm{Hz}$ is the laser cooling rate, $N = \langle {\hat{a}}^\dagger {\hat{a}} \rangle \approx 17$ is the steady-state occupation number, and the superoperator is defined by 
\begin{align}
\mathcal{D}_{\hat{c}} [\rho] = {\hat{c}} \rho {\hat{c}}^\dagger - \frac{1}{2}\{{\hat{c}}^\dagger {\hat{c}}, \rho \}.
\end{align}
Choosing length units such that $\sqrt{m \nu/\hbar} = 1$, this can be brought into our standard form \eqref{eq:QBM-general} with 
\begin{align}
\Hmat = \nu \Id, \quad \Dmat = \Gamma (N+1/2) \Id, \quad \gamma = \Gamma/2.
\end{align}
(Note that $\jdet{\Dmat} \ge \gamma^2$ since $N \ge 0$.)  Using Eq.~\eqref{eq:POVM-time} and the defining equation $\jdet{\Cmat_T} = (1+e^{-2 \gamma T})^2/4$ for $T$, we compute
\begin{gather}
\Cmat_t = (N+1/2) (1-e^{- t \Gamma}) \Id,\\
%T = \frac{1}{\gamma} \, \mathrm{ln} \left(\frac{2 \nu}{\Gamma(N+1/2)}\right),\\
T = \frac{2}{\Gamma} \, \mathrm{arccoth} \left(2N +1 \right),\\
%\CVTmat =\Cmat_T/(1+e^{-\Gamma T}).
\CVTmat = \IdT.
\end{gather}
The dynamics \eqref{eq:example-master} thus generate the evolution 
\begin{align}
\rho_0 \to \rho_T &= \left(\POVM{s}{\IdT} \circ \UCP_{\Rmat}\right) \left[\rho_0\right] %//
%&= \int \! \frac{\dd{\alvec}}{2\pi}  \ketbra{\sinv \alvec}{\alvec} \rho_0 e^{ i \hat{a}^\dagger \hat{a} T \nu} \ketbra{\alvec}{\sinv \alvec}
\end{align}
with 
\begin{gather}
s = e^{-\mathrm{arccoth} \left(2N +1 \right)}, \\
\Rmat = e^{\nu T \boldsymbol{\epsilon}} = \left( \begin{array}{cc} \cos \nu T & \sin \nu T \\ -\sin \nu T & \cos \nu T \end{array}\right).
\end{gather}
This is pure harmonic evolution for a time $T$ followed by a POVM measurement in the Gaussian basis of coherent states $\ket{\IdT;\alvec}$, where the corresponding prepared states are contracted toward the origin with proportionality constant $s \le 1$.

\end{document}